%% file: main.tex
\documentclass[conference]{IEEEtran}
\IEEEoverridecommandlockouts
\input{macros}
\def\BibTeX{{\rm B\kern-.05em{\sc i\kern-.025em b}\kern-.08em
    T\kern-.1667em\lower.7ex\hbox{E}\kern-.125emX}}
\begin{document}

\title{Evaluating Agent-based Program Repair at Google
}

\author{\IEEEauthorblockN{
Pat Rondon, Renyao Wei,
Jos{\'e} Cambronero, J{\"u}rgen Cito, 
Aaron Sun,\\ Siddhant Sanyam,
Michele Tufano,
Satish Chandra
}
\IEEEauthorblockA{
\textit{Google}\\
Email:\{rondon, renyaow, jcambronero, jcito, aasun, ssanyam, tufanomichele, chandrasatish\}@google.com
}
}

\maketitle

\input{abstract}


\input{introduction}
\input{gits-challenge-set}

\input{passerine}
\input{evaluating-agent-generated-repairs}

\input{results}

\input{discussion}

\input{threats}

\input{relatedwork}
\input{conclusion}

\bibliographystyle{IEEEtran}
\bibliography{references}

\end{document}

%% file: macros.tex
\usepackage[table]{xcolor}
\usepackage{etoolbox}
\usepackage{hyperref}
\usepackage{listings}
\usepackage{cite}
\usepackage{amsmath,amssymb,amsfonts}
\usepackage{algorithmic}
\usepackage{graphicx}
\usepackage{textcomp}
\usepackage{subcaption}
\usepackage{booktabs}
\usepackage{censor}
\usepackage[font=small,labelfont=bf]{caption}
\usepackage{xurl}

\usepackage{adjustbox}
\usepackage{xcolor}

\usepackage{multirow}

\newcommand{\cheapsubsection}[1]{\vspace{0.5em} \noindent{\textbf{#1}}}

%% file: abstract.tex
\begin{abstract}
Agent-based program repair offers to
automatically resolve complex bugs end-to-end by
combining the planning, tool use, and code generation
abilities of modern LLMs. Recent work has explored
the use of agent-based repair approaches on 
the popular open-source SWE-Bench~\cite{swebench}, a collection of bugs from highly-rated GitHub Python projects.
In addition, various agentic approaches such as SWE-Agent~\cite{sweagent} have been proposed to solve bugs in this benchmark.

This paper explores the viability of using an agentic approach to address bugs in an enterprise context. 
To investigate this, we curate
an evaluation set of 178 bugs drawn from
Google's issue tracking system. This dataset
spans both human-reported (78) and machine-reported bugs (100).

To establish a repair performance baseline on this benchmark, we implement Passerine, an agent similar in spirit to SWE-Agent that can work within Google's development 
environment.  We show that with 20 trajectory samples and Gemini 1.5 Pro,
Passerine can produce a patch that passes bug tests (i.e., plausible) for 
73\% of machine-reported and 25.6\% of human-reported bugs 
in our evaluation set. 
After manual examination, we found that 43\% of machine-reported bugs and 17.9\% of human-reported bugs have at least one patch that is semantically equivalent to the ground-truth patch.

These results establish a baseline on an industrially
relevant benchmark, which as we show,
contains bugs drawn from
a different distribution---in terms of language diversity, size, and spread of changes, etc.---compared to those
in the popular SWE-Bench dataset.

\end{abstract}

%% file: introduction.tex
\section{Introduction}\label{sec:intro}
Automated program repair (APR) has a long history in the programming languages and software engineering research communities. In the traditional setup, the APR system is given a bug-reproducing test suite and is tasked with fixing the bug. With the rise in machine learning methods, APR systems have increasingly relied on models (initially statistical and eventually deep learning-based) to perform critical APR tasks such as fault localization, patch generation, patch ranking and eventually end-to-end repair. More recently, work such as SWE-Agent~\cite{sweagent}, AutoCodeRover~\cite{autocoderover}, RepairAgent~\cite{repairagent},
CodeR~\cite{codeR},
AutoDev~\cite{autodev},
OpenDevin~\cite{opendevin}, and
others
have shown that, when incorporated into an agent-based system, LLMs can be used to perform end-to-end software engineering tasks in complex environments. Specifically, agentic repair systems can start from a bug description and autonomously generate bug-reproduction tests, localize faults, make candidate edits, validate these patches, and then submit a solution.

While such autonomous workflows have generated excitement in the APR community, systems in this space have been designed and evaluated using open-source bugs found in the GitHub ecosystem. In particular, SWE-Bench~\cite{swebench}, a collection of 2,294 Python bugs/fixes from popular GitHub repositories, and SWE-Bench-Lite, a subset of 300 bugs from SWE-Bench, have become the de-facto evaluation benchmarks for APR. %

It is not yet clear whether systems that perform well on SWE-Bench can achieve similar success when applied in the broader software industry, where we often encounter diverse collection of bugs which span a wide array of projects. Such conditions present both an opportunity and a challenge for APR systems. 
Given the enormous cost and effort to maintain code in enterprise environments, if agentic APR systems can perform as well in enterprise settings as on SWE-Bench, they hold the promise for substantial impact in industry.

To investigate the viability of agentic repair we first had to curate a benchmark set. Handling randomly-chosen bugs from Google's internal issue tracking system (GITS) would have been a non-starter for assessing agentic APR performance, due to various reasons, some of which are similar to those also encountered by the authors of SWE-Bench in their context: for example, each bug should have an easily-executable test.
Importantly, a randomly-chosen subset of bugs, while useful for population comparisons, provides little signal
for agent-level design and improvements.
Moreover, including bugs that
go beyond the current, but not future,
limitations of a basic agent
(e.g., screenshots) would
further cloud the informativeness
of any failures.

Consequently, we curated an evaluation set, GITS-Eval, of 178 bugs from Google's internal issue tracking system (GITS). 
This benchmark consists of 78 bugs reported by human developers and 100 machine-reported bugs, comprising 50 bugs reported by a suite of automated sanitizers (SAN), and 50 bugs reported by an automated test order dependency analyzer (TOD). These bugs reflect
different projects and programming languages, while remaining tractable for an APR system; the criteria for filtering are discussed in Section~\ref{sec:challengeset}.

To place GITS-Eval into context,  we also studied the differences
between SWE-Bench open source bugs and GITS. We sampled
2,000 bugs from GITS with filtering that reflects similar principles
to those underlying SWE-Bench and
found that GITS bugs exhibit different distributional characteristics from SWE-Bench open source bugs. In particular, we have observed differences in language diversity, size and spread of changes, and the presence of code terms in the bug description as a proxy for localization difficulty.
Specifically, we want to note that, due to these differences, the performance of an agent on one benchmark set may not be indicative of performance on the other.
In this paper, we make no claims on the performance of our baseline agentic system on SWE-Bench or SWE-Bench-Lite.

To establish a repair performance baseline on GITS-Eval, we built Passerine, a simple baseline agentic APR system inspired by SWE-Agent~\cite{sweagent} but designed to make use of Google’s internal development environment.
Using a limited command inventory, and with 20 sampled repair
runs (i.e. trajectories) per bug, Passerine is able to generate a
patch that passes bug tests (i.e. plausible) for 
25.6\% of the human-reported bugs, 73\% of machine-reported bugs (68\% of TOD-reported bugs and 78\% of SAN-reported bugs).
After manual annotation, we found that 17.9\% of human-reported bugs and 43\% of machine-reported bugs (24\% of TOD and 62\% of SAN) have at least one patch that was semantically equivalent to the ground-truth patch.
Thus, we establish that a simple agent can meaningfully solve issues from an industrially-relevant evaluation set. We expect
performance to improve as we
incorporate additional innovations
inspired by recent agentic
APR systems~\cite{specrover, autocoderover, codeR, marscode}.

We also make several observations about Passerine's behavior.
By analyzing command sequences, we note that Passerine adapts its behavior based
based on bug type (and associated bug information).
We found that Passerine can use rich bug reports, which carries implications for
bug report design in a future of increasingly prevalent agent-based systems.
Finally, we also found that trajectory analysis can reveal opportunities
for optimization, such as pruning degenerate trajectories.

The remainder of the paper is organized as follows.
Section~\ref{sec:challengeset} presents our approach to understanding the differences between SWE-Bench and GITS bugs and shows how we construct a representative challenge set -- GITS-Eval -- appropriate for measuring APR system performance at Google.
Section~\ref{sec:passerine} presents Passerine, a SWE-Agent inspired agentic repair system designed to operate within Google's development environment.
Section~\ref{sec:results} presents our main results: an evaluation of Passerine's performance on a challenge set of 178 GITS bugs and broader observations
for agent-based APR in an enterprise context.
Section~\ref{sec:threats} describes threats and limitations
to our work.
Section~\ref{sec:relatedwork} discusses related work in 
multiple areas of APR.

%% file: gits-challenge-set.tex
\section{Collecting a GITS Evaluation Set}
\label{sec:challengeset}

While SWE-Agent evaluated their system on SWE-Bench, a collection of open-source repository bugs and their associated fixes, an agent in the Google internal environment would face bugs of a different nature as a result of the environment's idiosyncrasies. To mention a few differences, Google uses a multilingual monorepo~\cite{potvingoogle}, with projects often spanning different portions of the repository, designed to build and run with Google-specific infrastructure (e.g., Bazel~\cite{bazeldistributed}), and with varying levels of domain-specific knowledge required to successfully develop in them. These differences have deep implications for not only writing, running, and testing code, but even for how the agent itself interacts with the Google internal environment (e.g. logging, which is sensitive business data and must be stored accordingly).

GITS, Google's internal issue tracking system, houses a vast and diverse collection of bugs spanning a wide array of projects. This presents both an opportunity and a challenge for automated program repair (APR) systems.
Opportunistically, randomly sampling bugs from GITS may seem appealing, as in principle the full database and constant stream of bugs could benefit from automated repair. 
However, properly selecting sensible bugs is challenging, as ad-hoc random sampling of bugs does not allow for repeatable progress measurement nor does it provide
an informative signal for the potential of APR.

\begin{figure}
    \centering
     \includegraphics[width=\columnwidth]{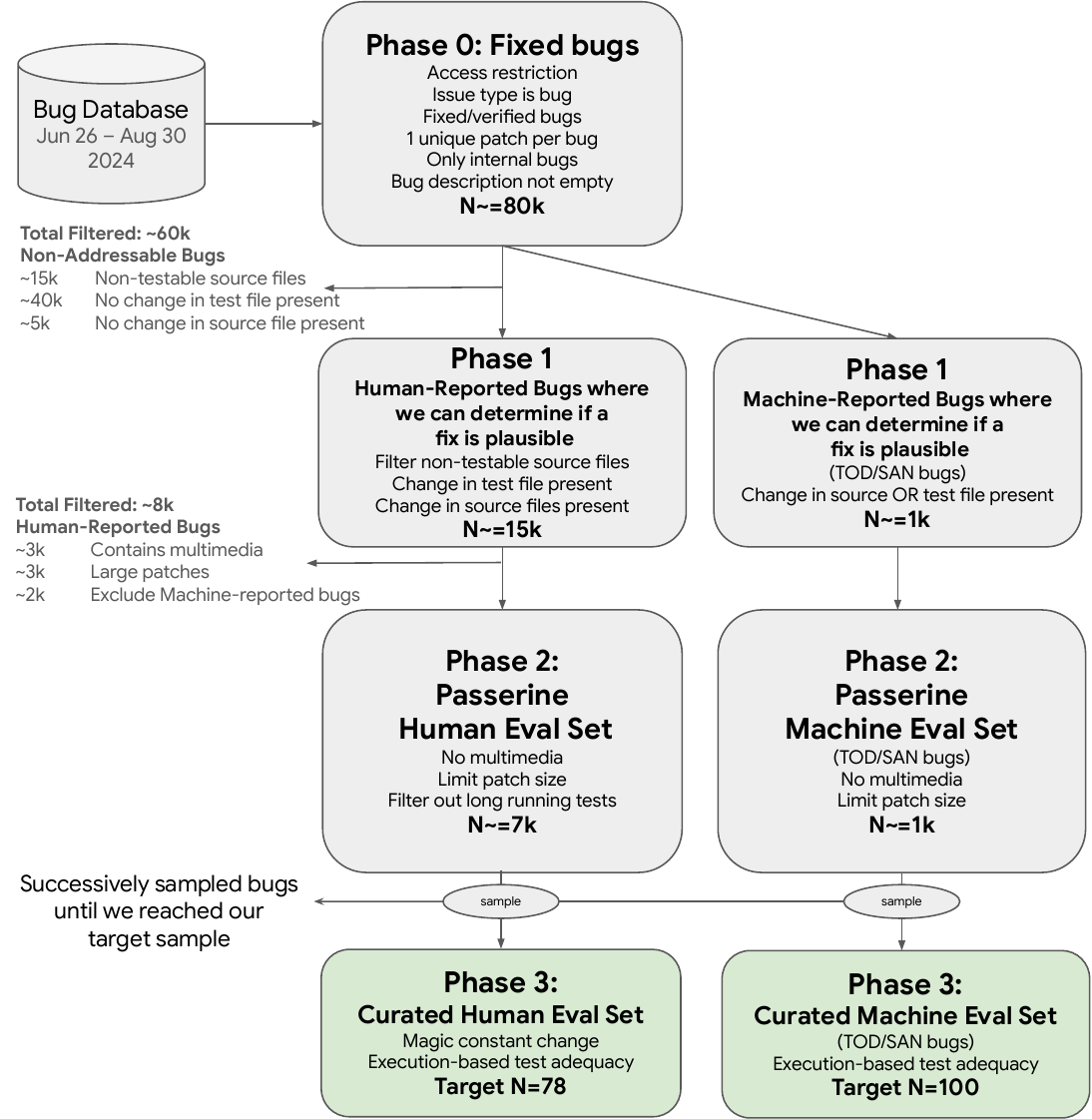}
    \caption{An overview of the different phases of filtering we go through to collect a GITS evaluation set}
    \label{fig:1p-population-curation-diagram}
\end{figure}

To effectively leverage this resource, we employ a multi-stage filtering funnel to curate a focused set of actionable bugs. This funnel ensures that the bugs presented to Passerine are both relevant to its capabilities and representative of real-world challenges within Google's codebase.
The filtering process comprises four distinct phases, as shown in Figure~\ref{fig:1p-population-curation-diagram}.

\subsection{Phase 0: Fixed Bugs Population}

This initial phase casts the widest net, aiming to define the broadest possible scope of relevant bugs. We apply minimal filtering criteria (Table~\ref{tab:population-definition-filters}), primarily to ensure accessibility and a clear association between bug reports and their corresponding fixes. This results in a large and diverse initial population of fixed bugs within a specific timeframe, serving as the foundation for further refinement.

\input{tables/population-definition-filters}

\subsection{Phase 1: Bugs Where We Can Determine if a Fix is Plausible
}
Phase 1 refines the selection by focusing on bugs that agentic APR could conceptually address, even if not currently supported by the agent's implementation, and for which we can assess whether the agent generated a plausible fix as part of the evaluation. This necessitates identifying bugs with testable code changes and verifiable fixes. A key aspect of this phase is the separation of human-reported bugs from machine-reported bugs. This distinction allows for tailored filtering criteria based on the nature of the bug report and the availability of information regarding the fix (Table~\ref{tab:phase1-filtering-criteria}).

\begin{figure}[h]
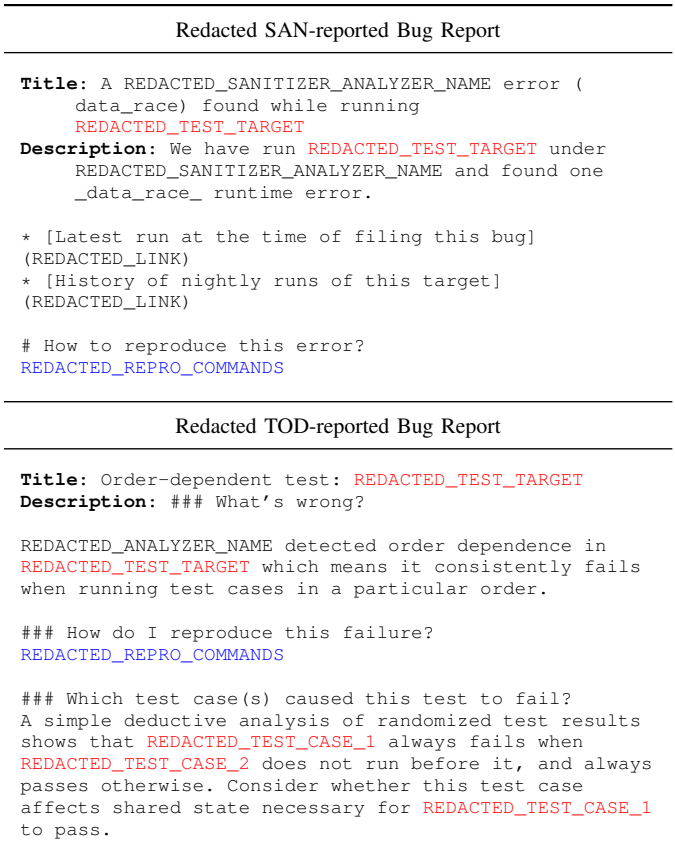

\vspace{-0.0cm}
    \centering
\begin{adjustbox}{width=0.49\textwidth}
\begin{tabular}{c}
\toprule
\footnotesize{Redacted SAN-reported Bug Report}\\
\midrule
\begin{minipage}[t]{0.45\textwidth}\vspace{-1em}
\begin{lstlisting}[escapeinside=||, breaklines]{text}
|\textbf{Title}|: A REDACTED_SANITIZER_ANALYZER_NAME error (data_race) found while running |\textcolor{red}{REDACTED\_TEST\_TARGET}|
|\textbf{Description}|: We have run |\textcolor{red}{REDACTED\_TEST\_TARGET}| under REDACTED_SANITIZER_ANALYZER_NAME and found one _data_race_ runtime error.

* [Latest run at the time of filing this bug]
(REDACTED_LINK)
* [History of nightly runs of this target]
(REDACTED_LINK)

# How to reproduce this error?
|\textcolor{blue}{REDACTED\_REPRO\_COMMANDS}|
\end{lstlisting}
\end{minipage}\\
\midrule
\footnotesize{Redacted TOD-reported Bug Report}\\
\midrule
\begin{minipage}[t]{0.45\textwidth}\vspace{-1em}
\begin{lstlisting}[escapeinside=||, breaklines=true]{text}
|\textbf{Title}|: Order-dependent test: |\textcolor{red}{REDACTED\_TEST\_TARGET}|
|\textbf{Description}|: ### What's wrong?

REDACTED_ANALYZER_NAME detected order dependence in 
|\textcolor{red}{REDACTED\_TEST\_TARGET}| which means it consistently fails
when running test cases in a particular order.

### How do I reproduce this failure?
|\textcolor{blue}{REDACTED\_REPRO\_COMMANDS}|

### Which test case(s) caused this test to fail?
A simple deductive analysis of randomized test results
shows that |\textcolor{red}{REDACTED\_TEST\_CASE\_1}| always fails when
|\textcolor{red}{REDACTED\_TEST\_CASE\_2}| does not run before it, and always
passes otherwise. Consider whether this test case
affects shared state necessary for |\textcolor{red}{REDACTED\_TEST\_CASE\_1}|
to pass.
\end{lstlisting}
\end{minipage}\\
\bottomrule
\end{tabular}
\end{adjustbox}
\vspace{-0.0cm}
\caption{Machine-reported bug reports, which typically contain richness such as reproduction information.}
\vspace{-0.0cm}
\label{fig:bug-reports}
\end{figure}

We focus on two types of machine-reported bugs (Figure~\ref{fig:bug-reports}): those surfaced by the SAN and TOD systems.
SAN performs a variety of sanitizer-based analyses, including memory and thread-related sanitizer reporting, which capture errors such as out-of-bounds accesses, uninitialized values, data races, and more.
Meanwhile, TOD automatically identifies test order-dependence. 

\input{tables/phase1-filtering-criteria}

\subsection{Phase 2: Automated Curation}
Phase 2 (Table~\ref{tab:phase2-filtering-criteria}) shifts the focus to practical considerations for evaluating our agent's current capabilities. Automated filters are applied to exclude bugs that would pose challenges for evaluation, such as those with long-running integration tests or requiring multi-modal understanding (e.g., screenshots in bug descriptions).

We limit the size of the patch in this phase to be less than 150 lines of code. This number represents the 90th percentile of bug fix patch sizes on a broad set of internal patches, ensuring that we are addressing a wide range of potential bugs.

\input{tables/phase2-filtering-criteria}

\subsection{Phase 3: Heuristic Curation}

To begin phase 3, we sample bugs from our phase 2 population and curate them incrementally.
Specifically, we execute any associated test with the
bug and ensure that there are appropriate failures 
before the ground-truth patch, which are then resolved
after the ground-truth patch is applied. In addition, to 
ensure consistent reproducibility, we remove
bugs that exhibit flaky behavior.

Finally, we introduce a layer of human expertise to ensure the quality and relevance of the benchmark set. We conduct manual reviews against a rubric to identify and filter out bugs that are not suited to automatic, execution-based evaluation, such as those relying on ``magic constants'' that cannot currently be easily captured by deterministic automated filters.
We define ``magic constants'' as either literal values (typically strings) or newly introduced code literals (such as method, class, or enum names) that appear in the updated test case but are not present in the original source code. Furthermore, these constants are not readily derivable from the bug report or any other existing code element. This characteristic implies that the successful execution of the bug confirmation test hinges on the precise value of this symbol or constant. However, a valid fix for the underlying bug may not necessitate the exact naming or literal value present in the ground-truth test.
To mitigate the potential for errors in this labeling process, at least one of the authors manually labeled each data point as containing a ``magic constant change'' or not, based on the definition provided above. In cases where the initial annotator was uncertain, a second author reviewed the data point to resolve any ambiguity.  A data point was only labeled as containing a "magic constant change" when both annotators concurred. This negotiated agreement approach helped ensure the reliability of our manual labeling process. 

For the future, we plan to explore automating these heuristics (e.g., through use of few-shot learning given our currently manually labeled set of examples).

This  multi-stage filtering process ensures that the resulting GITS evaluation set (GITS-Eval) is both representative of real-world bugs within Google's codebase and suitable for evaluating the capabilities and potential of agent-based APR. Our final evaluation set comprises 178 bugs: 78 human-reported bugs, and 100 machine-reported bugs, of
which 50 come from automated sanitizers (SAN) and 50 from an automated test order dependency analyzer (TOD).

\subsection{GITS vs SWE-Bench Bugs}
Understanding the nature of our evaluation set is crucial for interpreting the results of our analysis. To provide context, we compare the distribution of GITS bugs with those in the GitHub-derived SWE-Bench dataset, highlighting important differences in their distributions.
While SWE-Bench does not reflect the entirety of GitHub, it is a widely-used benchmark which sets the standard for evaluating automatic program repair systems.
Thus, to create a comparable test set for GITS, we draw a random sample of 2,000 bug-fixing patches from Phase 1 of our bug filtering phases which comprise both human- and machine-reported bugs for which we can determine a plausible fix. We compare SWE-Bench to Phase 1 bugs, rather than later manually-curated bugs, to capture intrinsic differences in bug distribution without confounding this comparison as a result of Passerine's current limitations (e.g. no multimedia, no flaky tests) or infrastructure challenges (e.g. long running tests).%
This is roughly in line with the methodology used for curation of the SWE-Bench dataset.

We will compare distribution differences along dimensions that correlate with localization difficulty and editing difficulty to highlight the unique challenges Google internal bugs need to address.

\subsubsection{Localization}

Before a candidate patch can be generated, a fault must be localized to identify what program statements are the root cause for the bug and so should be modified to produce a valid fix. While the difficulty of localization can be influenced by many factors, we consider two aspects: searchability and spatial distribution of changes. 

\begin{figure*}[t!]
    \centering
    \begin{subfigure}[b]{0.19\textwidth}
        \centering
        \includegraphics[width=\textwidth]{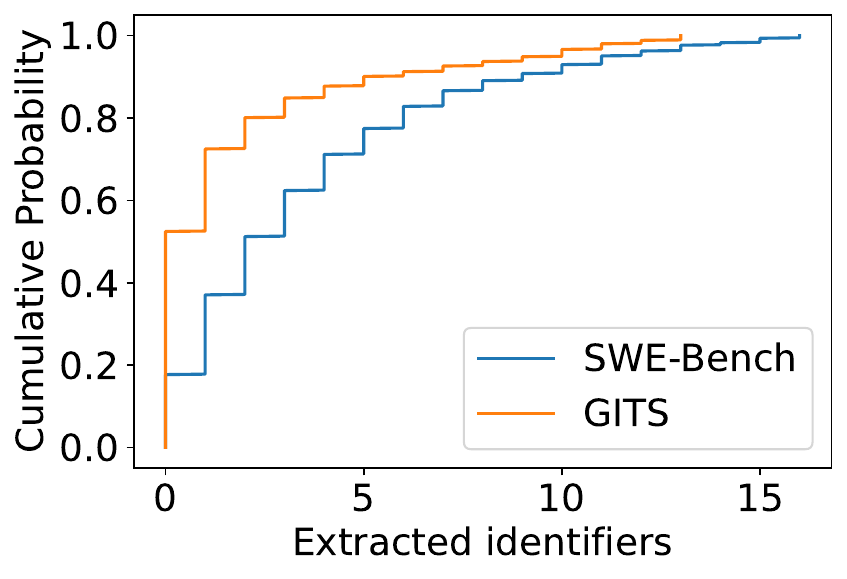}
        \caption{}
        \label{fig:rq1-localization-proxy}
    \end{subfigure}
    \hfill    
    \begin{subfigure}[b]{0.19\textwidth}   
        \centering
        \includegraphics[width=\textwidth]{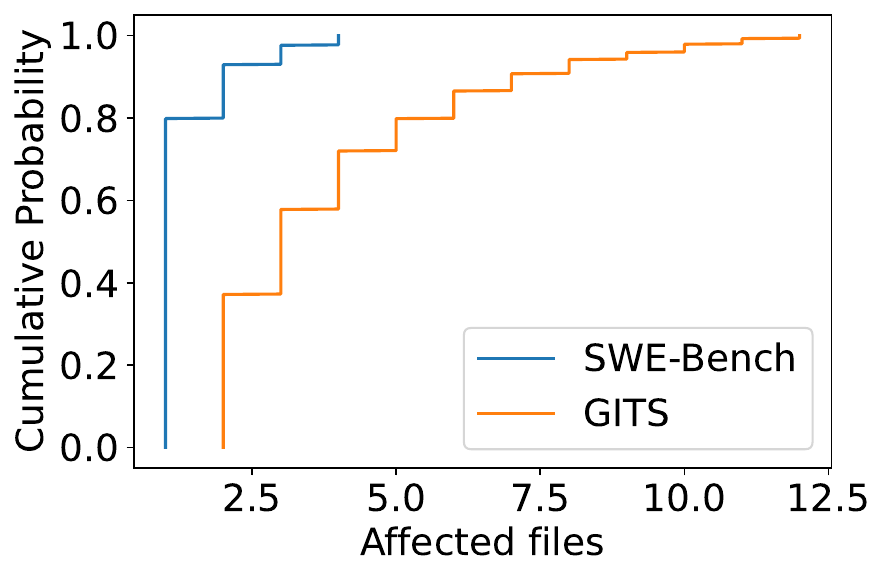}
        \caption{}
        \label{fig:spread_files}
    \end{subfigure}
    \hfill
    \begin{subfigure}[b]{0.19\textwidth}
        \centering
        \includegraphics[width=\textwidth]{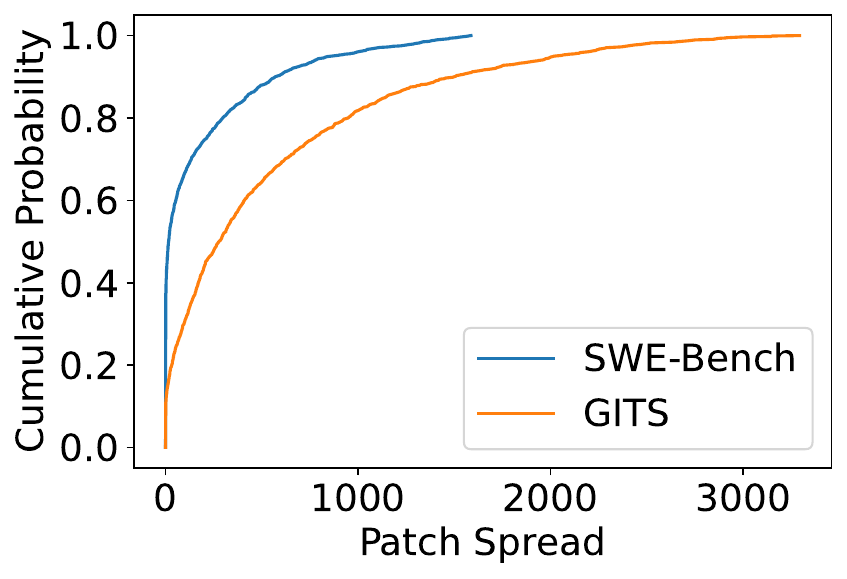}
        \caption{}
        \label{fig:spread_lines}
    \end{subfigure}
    \hfill
    \begin{subfigure}[b]{0.19\textwidth}
        \centering
        \includegraphics[width=\textwidth]{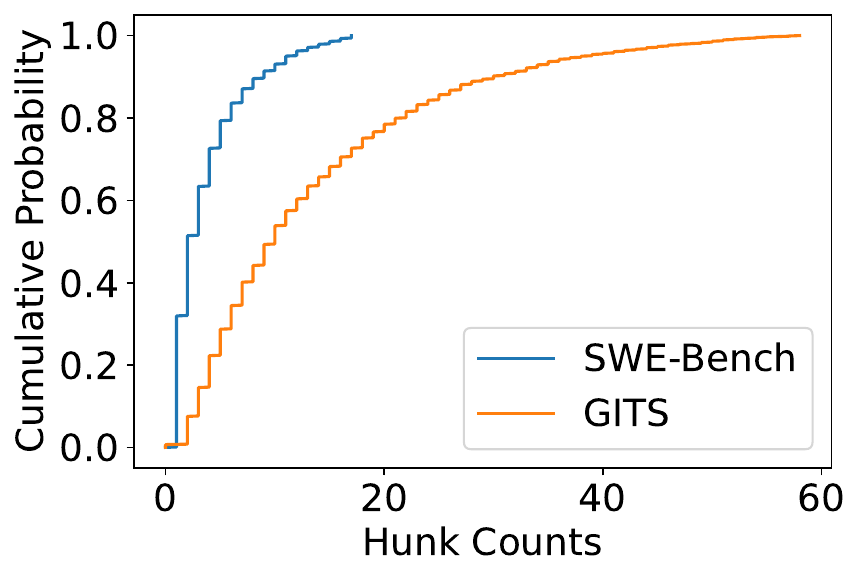}
        \caption{}
        \label{fig:spread_hunks}
    \end{subfigure}
    \hfill
    \begin{subfigure}[b]{0.19\textwidth}
        \centering
        \includegraphics[width=\textwidth]{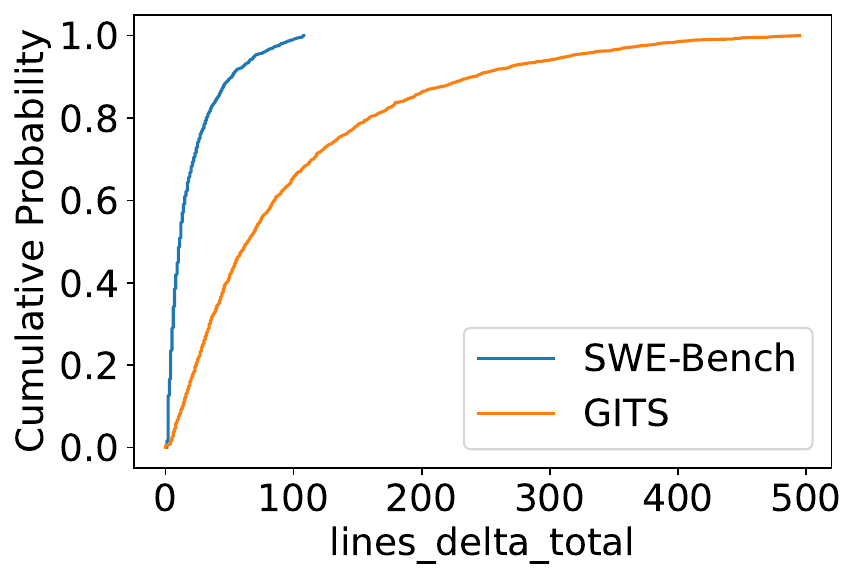}
        \caption{}
        \label{fig:rq1-lines-delta}
    \end{subfigure}    
    \caption{Comparing GITS Phase 1 bugs to SWE-Bench. Changes found in ground-truth patches solving GITS bugs tend to (a) have fewer likely code identifier tokens, which in turn require more sophisticated code search abilities and bug knowledge to successfully localize the original fault. These patches modify substantially more files (b), which are further apart in the codebase (c), and produce changes with many separate hunks (d)
    and more lines of code changed (e).
    }
\end{figure*}

Code search typically plays a substantial role in localization, as it allows a user to navigate large and potentially unfamiliar codebases~\cite{sadowski2015developers}. Often, code search starts by identifying terms that are relevant to the bug and that may appear in the underlying codebase. To emulate this process, we consider the frequency of terms in a bug issue description that are likely to be codebase symbols, such as class names. We extract these terms using a simple regex-based heuristic, where any term that matches either \texttt{snake\_case} or \texttt{CamelCase} identifiers is considered a code term.
Figure~\ref{fig:rq1-localization-proxy} shows an empirical cumulative distribution function (ECDF) of the number of possible code terms in the associated bug descriptions. We find that GITS bugs have fewer possible code terms in their descriptions.
Only 18\% of GITS bugs have at least 2 possible code terms, compared to approximately 60\% in SWE-Bench.

Next, we consider the extent to which fixes are spatially related. We first compare the number of different files modified by a patch, as well as the number of hunks resulting from the segmentation of those changes~\cite{hunks}. Finally, we also consider patch spread, defined as the number of lines of separation between continuous hunks~\cite{sobreira2018dissection}, as a measure of how much a patch is dispersed throughout a file or across multiple files.
We compute the number of unmodified lines between consecutive changes for each affected file and sum these to yield the patch spread. A higher patch spread suggests a greater degree of interleaving, where modified lines are scattered throughout the file. Conversely, a lower spread indicates that modifications are clustered together in contiguous blocks.

Figures~\ref{fig:spread_files} and~\ref{fig:spread_lines} show that GITS patches modify more files (up to twice as many), and that these modifications result in much larger hunk counts, respectively. When we measure patch spread (Figure~\ref{fig:spread_lines}), we again find that patches are more widely separated within files than those in SWE-Bench.

\subsubsection{Editing}

Once a fault location is identified, the associated statements must be edited to produce a fix. Intuitively, larger edits (meaning more lines of code) can pose a challenge as they introduce more opportunities for mistakes. We measured the number of lines changed in ground-truth patches. As shown in Figure~\ref{fig:rq1-lines-delta},
almost all patches from SWE-Bench are under 100 lines while
only
approximately 40\% of GITS patches are under 100 lines.

This complexity is further compounded by the need to consider the specific syntax and semantics of different programming languages. To better understand the language distribution we identified the top 5 most frequent file extensions in our sample of GITS patches (SWE-Bench only considers Python files): Java, C++, TypeScript, Kotlin, and Python.

\vspace{-0.4cm}
\begin{center} 
\fbox{
\begin{minipage}[t]{0.97\linewidth}
{\bf GITS vs SWE-Bench.} 
Our analysis shows that GITS bugs present unique challenges compared to SWE-Bench, particularly in terms of localization and code modification. GITS and SWE-Bench differ slightly in the amount of code-related terms in their descriptions. Additionally, patches for GITS bugs differ in nature when it comes to  changes across files, the number of modified lines, and dispersion of the modifications (i.e., patch spread).
\end{minipage}
}
\end{center}

\subsection{GITS-Eval vs. SWE-Bench-Lite}
We previously introduced GITS-Eval, a more tractable, curated subset of 178 bugs, which we use for evaluation of our agentic repair system.
In an open source context, most state-of-the-art agent-based APR approaches are evaluated on SWE-Bench-Lite, a more practical and self-contained benchmark derived from SWE-Bench. GITS-Eval is the analogous dataset in our context.
Recognizing the potential for variations between human-reported and machine-reported bugs, we have categorized the bugs within GITS-Eval accordingly. 

\begin{figure*}[t!]
    \centering
    \begin{subfigure}[b]{0.19\textwidth}
        \centering
        \includegraphics[width=\textwidth]{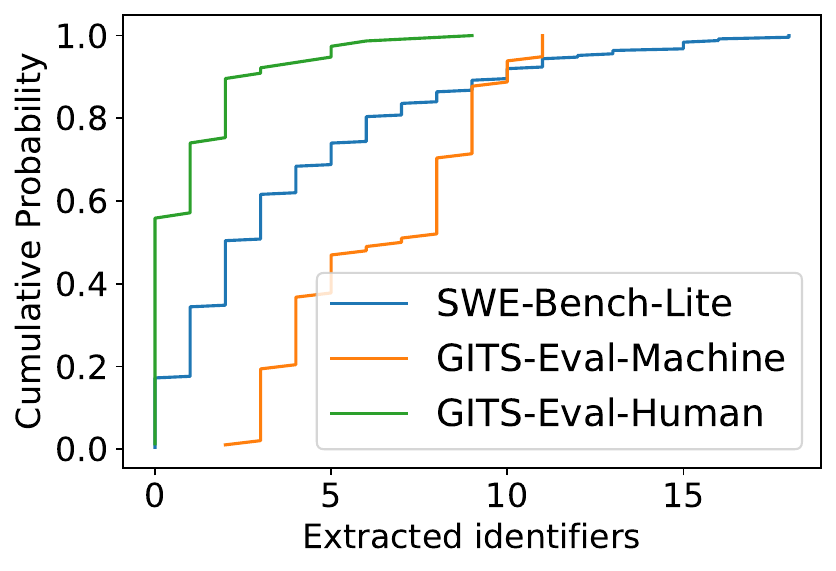}
        \caption{}
        \label{fig:localization_eval}
    \end{subfigure}
    \hfill    
    \begin{subfigure}[b]{0.19\textwidth}
        \centering
        \includegraphics[width=\textwidth]{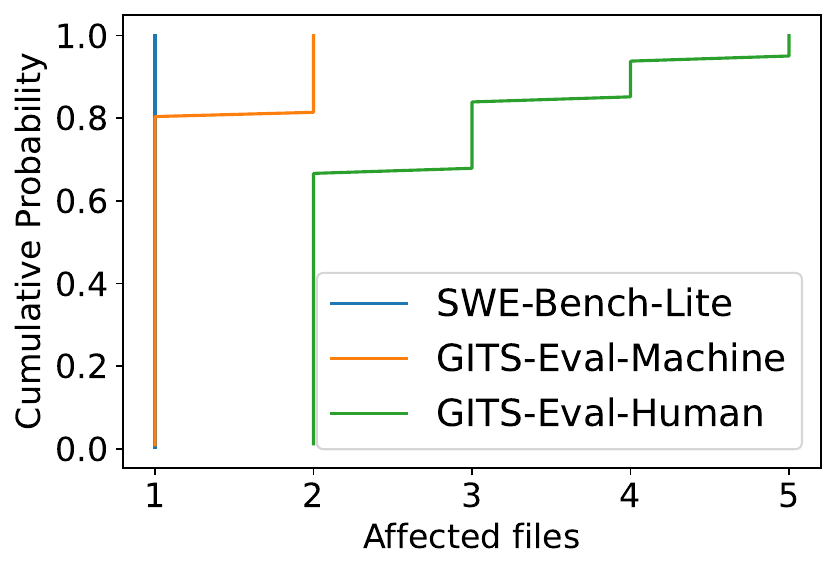}
        \caption{}
        \label{fig:spread_files_eval}
    \end{subfigure}
    \hfill
    \begin{subfigure}[b]{0.19\textwidth}
        \centering
        \includegraphics[width=\textwidth]{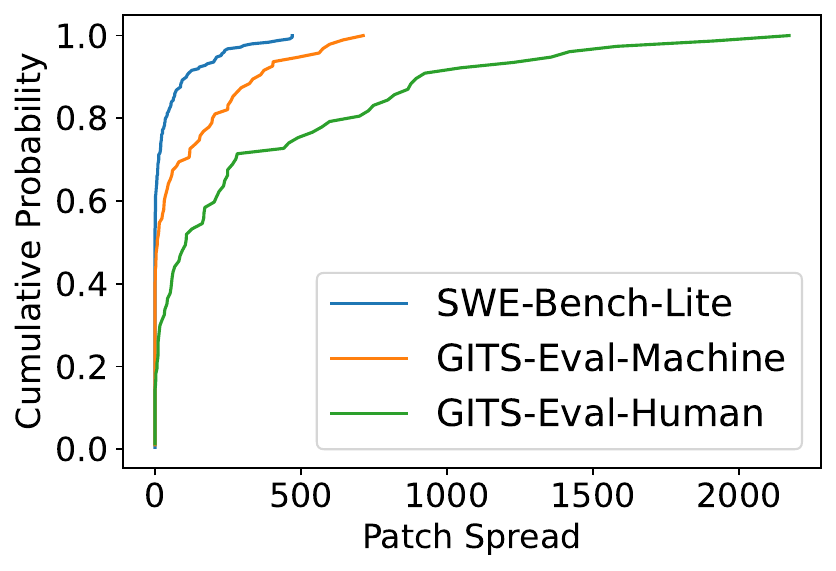}
        \caption{}
        \label{fig:spread_lines_eval}
    \end{subfigure}
    \hfill    
    \begin{subfigure}[b]{0.19\textwidth}
        \centering
        \includegraphics[width=\textwidth]{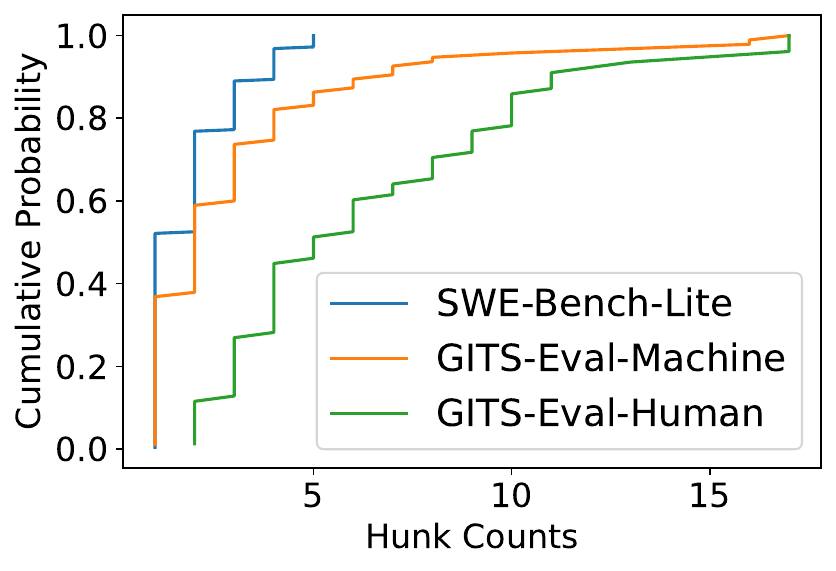}
        \caption{}
        \label{fig:spread_hunks_eval}
    \end{subfigure}
    \hfill
    \begin{subfigure}[b]{0.19\textwidth}
        \centering
        \includegraphics[width=\textwidth]{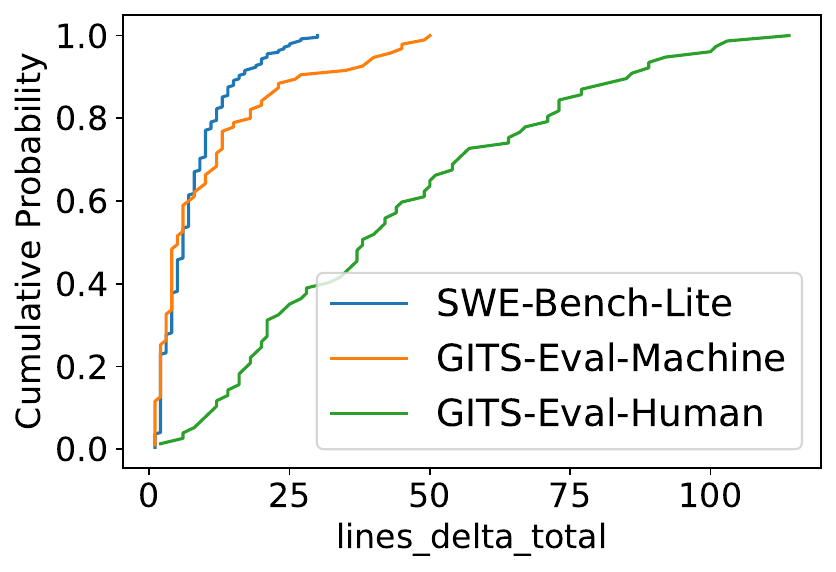}
        \caption{}
        \label{fig:rq1-lines-delta_eval}
    \end{subfigure}    
    \caption{
    GITS-Eval machine-reported bugs are comparable to SWE-Bench-Lite, while human-reported 
    GITS-Eval bugs are more complex and resemble differences in our GITS and
    SWE-Bench comparison.
    }
    \label{fig:rq1-eval}
\end{figure*}

Figure~\ref{fig:rq1-eval} compares SWE-Bench-Lite, human, and machine-reported
bugs in GITS-Eval. We find that machine-reported bugs in GITS-Eval are comparable to bugs
in SWE-Bench-Lite. Meanwhile, human-reported GITS-Eval bugs display
increased complexity compared to SWE-Bench-Lite, resembling the differences
observed in our earlier comparison of GITS and SWE-Bench.

%% file: tables/population-definition-filters.tex
\begin{table}[]
\resizebox{\columnwidth}{!}{
    \rowcolors{2}{gray!25}{white}
    \begin{tabular}{ll}
        \rowcolor{gray!50}
    \hline
    Filtering Criteria  & Justification                                                                                                                             \\ \hline
    Access Restriction  & Bug and associated fix must be accessible for analysis.                                                                                   \\
    Issue Type is Bug   & Must be classified as a ``bug'' in our bug database.                                                                                        \\
    Bug Status          & Bug status is ``fixed''.                                                                                                                    \\
    Patch Association   & One bug is associated with one patch, and vice versa.                                                                                     \\
    Bug Description     & Must have a non-empty description                                                                                                         \\
    Project Scope       & \begin{tabular}[c]{@{}l@{}}Restricted to the main Google repository\\
                        (Passerine's current scope).\end{tabular}                                                                     \\
    Bug Source          & Excludes externally reported bugs.                                                                                                         \\
    Patch Data Range    & \begin{tabular}[c]{@{}l@{}}Fix patches are sampled from June 26 to August 30, 2024\\ (chosen for repeatability and to avoid data contamination).\end{tabular} \\
    Bug Creation Cutoff & \begin{tabular}[c]{@{}l@{}}Bugs created no earlier than one year\\
                                before the patch date range start.\end{tabular}                                                                  \\ \hline
    \end{tabular}
}
\caption{Phase 0 
Filters}
\label{tab:population-definition-filters}
\end{table}

%% file: tables/phase1-filtering-criteria.tex
\begin{table}[]
\resizebox{\columnwidth}{!}{
    \begin{tabular}{ll}
    \rowcolor{gray!50}
    \hline
    Filtering Criteria     & Justification                                                                                                                    \\ \hline
    Testable source files  & \begin{tabular}[c]{@{}l@{}}Excludes patches affecting file types that are difficult to \\ test (e.g., sql, html, css, config files).\end{tabular}
    \\ \hline
    \rowcolor{gray!50}
    \multicolumn{2}{l}{Filtering Criteria for Human-Reported Bugs}                                                                                            \\ \hline
    Change in test file    & \begin{tabular}[c]{@{}l@{}}Requires at least one test file to be affected by the patch.\end{tabular}                          \\
    \rowcolor{gray!25}
    Changes in source file & \begin{tabular}[c]{@{}l@{}}Requires at least one source code file (e.g., .cc, .java, .py,\\ .dart, .js, .ts, .go) to be affected by the patch.\end{tabular}
    \\ \hline
    \rowcolor{gray!50}
    \multicolumn{2}{l}{Filtering Criteria for Machine-Reported Bugs}                                                                                          \\ \hline
    Change in code file    & \begin{tabular}[c]{@{}l@{}}The change affects at least a code file that could be either \\ a source code file or a test file.\end{tabular}
    \\ \hline    
    \end{tabular}
}
\caption{Phase 1 Filters}
\label{tab:phase1-filtering-criteria}
\end{table}

%% file: tables/phase2-filtering-criteria.tex
\begin{table}[]
\resizebox{\columnwidth}{!}{
    \begin{tabular}{llc}
    \rowcolor{gray!50}
    \hline
    Filtering Criteria & Justification & Phase
    
    \\ \hline
    Patch size limit & \begin{tabular}[c]{@{}l@{}}Limits patch sizes to \textless{}150 lines of code for tractability.\end{tabular} & 2      
    \\ \rowcolor{gray!25}
    No multimedia & \begin{tabular}[c]{@{}l@{}}Excludes bugs with multimedia  content in the description.\end{tabular} & 2
    \\
    \begin{tabular}[c]{@{}l@{}}Execution-based \\ Test Adequacy\end{tabular} & \begin{tabular}[c]{@{}l@{}}Confirms that tests associated with the fix reveal the bug, \\ are not flaky, and validate the fix.\end{tabular} & 3                                   
    \\ \rowcolor{gray!25}
    \begin{tabular}[c]{@{}l@{}}No magic constant  \end{tabular} & \begin{tabular}[c]{@{}l@{}}Excludes bugs where fix requires new literals or code symbols\\that cannot be inferred from context.\end{tabular} & 3 \\ \hline                 

    \rowcolor{gray!50}
    \multicolumn{3}{l}{Filtering Criteria for Human-Reported Bugs}                                                                                            \\ \hline
    Long-running tests & \begin{tabular}[c]{@{}l@{}}Excludes bugs with links to our internal integration \\ test platform (likely indicating long test runs).\end{tabular} & 2 \\ \hline
    \end{tabular}
}
\caption{Phase 2 and 3 Filters}
\label{tab:phase2-filtering-criteria}
\end{table}

%% file: passerine.tex
\section{Passerine: An Agent-based Repair System}
\label{sec:passerine}

To establish a repair performance floor for agent-based APR on GITS-Eval, we developed an APR agent, Passerine, inspired by SWE-Agent.
Like SWE-Agent, Passerine uses a ReAct~\cite{react} loop to iteratively produce ``thoughts,'' run commands against the current state of the workspace, and observe the commands' results; the output of the agent is a modified workspace in which, if the agent is successful, the bug is fixed, as well as a trace of the agent's execution, for human interpretation and debugging.
Similarly to SWE-Agent, we expose commands to the agent that are suited both to the APR task and designed with the ergonomics of the ``agent-computer interface'' in mind; these commands include both workalikes of the SWE-Agent commands (file viewing, file editing, and terminating the agent), as well as commands that provide functionality specific to Google's development environment (compiling and running tests, searching the index of Google's monorepo).

\emph{Unlike} SWE-Agent, we find that we do not need to expose a full Linux command line to the agent, e.g., via containerization.
Google's test-running infrastructure already provides sufficient containerization~\cite{potvingoogle}. Our experience, as well as prior work in agent-based APR like RepairAgent~\cite{repairagent} and AutoCodeRover~\cite{autocoderover}, shows that a small, APR-focused command set is sufficient for bug-fixing.

Passerine is intentionally simple and minimal, avoiding complex architecture or explicitly dividing the APR process into discrete phases like localization, editing, testing, and so forth.
This simple approach is validated by the effectiveness of agents without prespecified control flow like SWE-Agent~\cite{sweagent}, CodeAct~\cite{codeact}, and OpenDevin~\cite{opendevin}.
We show, through the first evaluation of a simple SWE-Agent-like APR system in an industrial setting, that, in spite of its minimalism, our agent is capable of addressing many real-world bugs.

\subsection{Agent Design}

\begin{figure}
    \centering
    \includegraphics[width=\columnwidth]{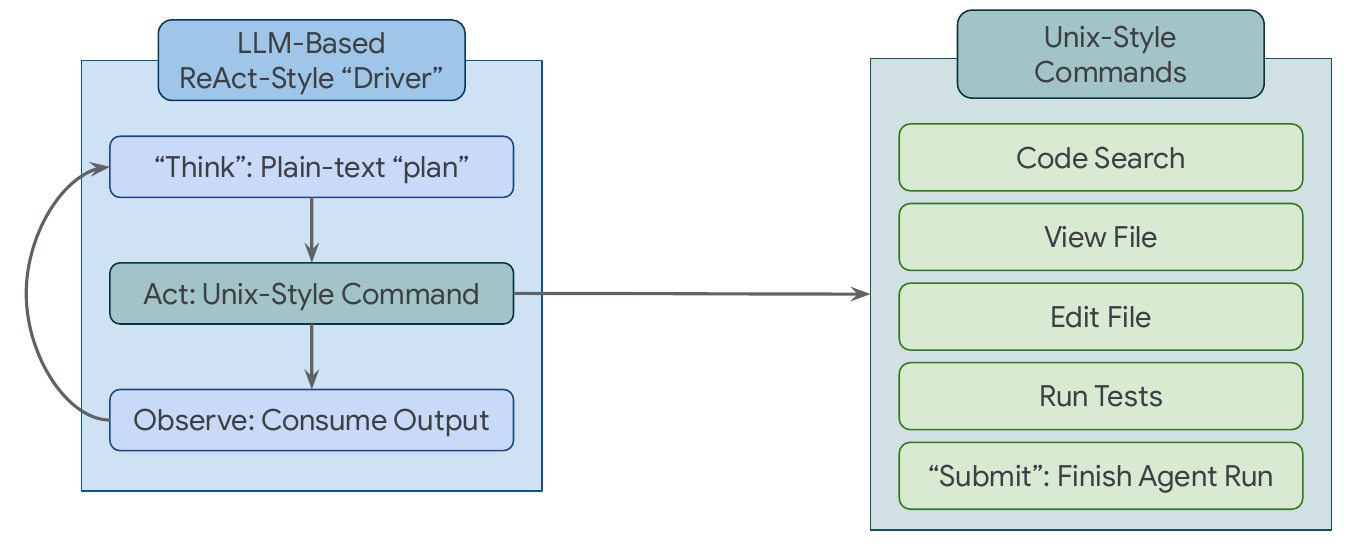}
    \caption{A high-level overview of Passerine’s ReAct-style dynamic loop and the Unix-style (adapted to Google) commands exposed as tools for the agent to use.
    }
    \label{fig:passerine-high-level-react}
\end{figure}

Passerine is a fully ``dynamic'' agent in the sense that it has no prespecified control flow. Every step in Passerine corresponds to a ReAct~\cite{react} style step, depicted
in Figure~\ref{fig:passerine-high-level-react},
where the agent issues an LLM prompt asking for the next command to perform and obtains back a response in the form of “thought” and “action.” The thought is a natural language fragment reasoning about the state of the agent and describing the next step. The action is a code fragment consisting of a Unix-like tool name, associated arguments for its call, and, for some commands (e.g., text editing), multiline input text. The agent framework then parses that command, executes it,
and replies to the agent with any observable outcomes.

\cheapsubsection{LLM Prompting and History:} When querying the LLM to obtain a new step, the agent includes information on the current state of the environment. Currently, 
Passerine includes the entire history of the agent. This has the benefit of simplicity and is facilitated by longer-context models. However, as we describe in Section~\ref{sec:challenges}, we can still encounter agent runs that exceed LLM context limits and thus we could benefit from more-sophisticated history management strategies.

\cheapsubsection{Agent Termination:} Passerine can call a \lstinline!finish [success|failure]! command to terminate, along
with the agent's judgment of the outcome.
To avoid non-termination, we also enforce a maximum step limit.

\cheapsubsection{Isolation:} Because the agent makes stateful changes to the environment, we rely 
on Google's codebase containerization~\cite{potvingoogle} to enforce isolation between different agent executions. We describe more of the infrastructure associated with running the agent in Section~\ref{sec:evaluation-framework}.

\subsection{Commands}\label{sec:commands}
Passerine takes a minimalist approach by providing a set of only 5
commands (and 1 additional alias commonly referenced in bug descriptions) that the agent can use, which are designed to interact with
Google internal APIs. A direct benefit of using a limited command set, restricted only to a few operations that can be implemented on top of already-containerized APIs,
is that Passerine \emph{does not} require a full virtual machine for isolation, improving scalability.

Commands in our design have a uniform input and output interface. For inputs, commands accept zero or more positional arguments, followed by arbitrary text fragments after a newline (e.g. to support editing). Every command output consists of an exit code and output text. The output text is the observable behavior that is exposed to the agent in its history.

We now briefly describe each command shown on the right-hand side of Figure~\ref{fig:passerine-high-level-react}.

\begin{itemize}
    \item \textbf{cat:} Takes a single file in the workspace and lists its contents using internal APIs. Every line is prefixed with a line number.
    
    \item \textbf{code\_search}: Takes one or more terms and concatenates these to issue a query for the Google internal Code Search API~\cite{sadowski2015developers}.

    \item \textbf{edit:} Takes a filename, starting and ending lines, and the replacement text with which to substitute the identified line region. When the edit is applied, the command's output text corresponds to the modified content (along with up to 3 context lines around it), where lines have been prefixed with the resulting line numbering post-edit. Edits are done through an internal API, which interacts with the isolated codebase container.

    \item \textbf{bazel}: Takes a test suite target, along with arbitrary parameters for test suite execution, and issues a distributed build/execution~\cite{greenberg2016building} for the associated target  using Google's internal build tool Bazel~\cite{wang2021smart}. The framework then parses the build report, extracts outcomes and relevant log excerpts, and surfaces these as an observable output for the agent.

    \item \textbf{finish}: Takes as argument 
    ``\lstinline|success|'' or ``\lstinline|failure|''
    and terminates agent execution.

\end{itemize}

All commands are implemented as simple Python functions, which easily support incorporating command validation and logging logic. We present each command to Passerine in its
initial prompt, providing a high-level description of the behavior, along with
fixed basic examples showing how each command can be used. To illustrate, Figure~\ref{fig:prompt-skeleton} shows our agent's system prompt template, along with a command doc template which is populated by each command's implementation.

\begin{figure}[h]
\vspace{-0.0cm}
    \centering
\begin{adjustbox}{width=0.49\textwidth}
\begin{tabular}{c}
\toprule
\footnotesize{System prompt template}\\
\midrule
\begin{minipage}[t]{0.45\textwidth}\vspace{-1em}
\begin{lstlisting}[escapeinside=||, breaklines]{text}
SETTING: You are a software engineer working on fixing bugs.

You are only allowed to use the following commands:

COMMANDS:
LIST_OF_COMMAND_DOCS

Output instructions:
1. Your output should always include only one thought section and one action.
2. Explain what you are doing in the thought section.
3. An action should be a single command from the COMMANDS section.
4. Do not include more than one command in the action.

Here is an example output:
THOUGHT:
First I'll start by running the test to reproduce the issue.
ACTION:
```tool_code
bazel test //some/target
```
\end{lstlisting}
\end{minipage}\\
\midrule
\footnotesize{Command doc template}\\
\midrule
\begin{minipage}[t]{0.45\textwidth}\vspace{-1em}
\begin{lstlisting}[escapeinside=||, breaklines]{text}
Command name: COMMAND_NAME
# COMMAND_NAME
## Description:
DESCRIPTION_TEXT
## Usage:
USAGE_TEXT
## Examples:
EXAMPLE_TEXT
\end{lstlisting}
\end{minipage}\\
\bottomrule
\end{tabular}
\end{adjustbox}
\vspace{-0.0cm}
\caption{System and command documentation template for Passerine}
\vspace{-0.0cm}
\label{fig:prompt-skeleton}
\end{figure}

\subsection{Evaluation Framework for Historic Bugs}\label{sec:evaluation-framework}

As part of developing Passerine, we also implement a framework which allows us to evaluate different agent configurations on Google infrastructure and analyze their performance, all in the context of historical bugs that have already fixed.

\cheapsubsection{Setup:} Before evaluating an agent, the framework sets up the environment by 1) loading the appropriate bug information and 2) checking out the repository state (also known as a changelist, which is the snapshot that undergoes code review\footnote{Google engineering practices documentation: \url{https://github.com/google/eng-practices}}) \emph{prior} to the associated ground-truth fix. Then
the framework reproduces the bug in this environment, to confirm that the setup correctly exposes the issue as expected. To do so, our framework allows benchmark repair tasks to specify the expected test targets to run and any files that are relevant to that state, and may need to be copied over from the ground-truth patch.
In practice, for machine-reported bugs, our framework automatically extracts bug-reproducing tests from the associated issue content using regular expressions.
Note that, for TOD tests, the associated test specifies a \emph{single} ordering of the test cases which exhibits the order dependency.
For human-reported bugs,
our framework has an offline pass that uses Google
tooling to identify test targets that depend on the test
files modified by the patch, considers these as candidate 
bug reproducing tests, and prunes the set down to those
that fail before and pass after the ground-truth patch has
been applied.

After the bug has been confirmed, the framework reverts the repository state to reflect only information available before the ground-truth fix was produced.

\cheapsubsection{Execution:} During execution, the agent operates independently, but the framework records agent-related events into a structured log for offline analysis. Logs reflect events such as commands issued by the agent and outputs obtained from the environment, as well as information such as what files/lines have entered the agent history (for localization analysis).
Importantly, our evaluation framework ensures, that during
command execution, there is no data leakage from future
states in the repository, e.g. code search results
are limited to those up to the commit preceding the ground-truth fix.

\cheapsubsection{Cleanup:}
After the agent has finished its execution (either through step budget exhaustion or issuing a finish command), the framework once again patches in any necessary test files and executes the bug reproduction test targets to confirm that the agent’s changes to the repository state result in a successful execution. 

(Note that \emph{specifically} the ground-truth test file is used to evaluate the plausibility of agent fixes for human-reported bugs; also, we do not evaluate agent modifications to test files, leaving this to future work.)

At this point, the framework also records structured logs of the execution, as well as evaluation reproduction information.

%% file: evaluating-agent-generated-repairs.tex
\section{Evaluating Agent-Generated Repairs}\label{sec:metrics}

We evaluate Passerine on GITS-Eval considering dimensions of overall patch correctness and trajectory dynamics.

\cheapsubsection{Patch Correctness} Consistent with prior APR work, we distinguish
between \emph{plausible} and \emph{valid} patches. In our work, a \emph{plausible} patch is
defined as patch that enables successful execution of the bug-reproducing test suite
associated with the repair task. A \emph{valid} patch~\cite{plausible1, plausible2, plausible3}, in turn, 
is defined as patch that implements a fix that is semantically equivalent to the one in the ground-truth patch~\cite{efficiencyofrepair,laregscalecorrectness}. In cases where ground-truth patches include additional actions like adding or modifying tests, we focus solely on whether the plausible patch addresses the bug-fixing logic.  (Generating regression tests is beyond the scope of this evaluation.)
We determine this correctness through manual analysis performed by three of the authors. During the process, one reviewer analyzed each patch, consulting with the other two authors for assessment of complex cases.
In a practical deployment of Passerine, patches could additionally be tested using Google's internal continuous integration platform, which runs extensive project-specific tests. However,
for our evaluation we focus on manual grading of validity, as  project-specific tests, like all testing, are generally insufficient for establishing or rejecting equivalence.
Note that we do not evaluate the textual similarity of the generated patches to the ground truths.

\cheapsubsection{Trajectory Dynamics} We analyze and compare Passerine's repair trajectories across several dimensions:
\begin{itemize}

    \item \textit{Trajectory Strategies}: We investigate the types and sequences of commands
    that Passerine employs during the repair process. This helps us understand how the agent approaches different bug types and whether there are distinct patterns in its actions.
    
    \item \textit{Localization}: We examine how effectively Passerine pinpoints the correct file(s) to modify by measuring the filesystem distance between the files modified by Passerine and the actual location of the ground truth bug fix.
    
    \item \textit{Trajectory Smells}: We analyze the presence of potentially suboptimal or unusual patterns within the trajectories, which we refer to as ``trajectory smells,'' inspired by code smells~\cite{fowler2018refactoring}. We consider four specific ``smells'': 
    (i) \texttt{NO\_TEST\_SMELL}: trajectory does not include any test execution commands (i.e., agent never confirmed the bug nor tested the patch);
    (ii) \texttt{NO\_OP\_CAT\_SMELL}: trajectory contains instances where a file is re-read without any intervening modifications to that file (i.e., agent made an unnecessary read);
    (iii) \texttt{CONSECUTIVE\_SEARCH}: trajectory contains
    at least three code search commands in a sequence (i.e., agent is repeatedly searching);
    (iv) \texttt{CONSECUTIVE\_EDITS}: trajectory contains
    at least 3 edits to the same file in a sequence (i.e., agent is repeatedly editing the same file);
\end{itemize}

%% file: results.tex
\section{Results}
\label{sec:results}

We present the results on our evaluation set of 178 GITS-Eval bugs, comprising 50 TOD bugs, 50 SAN bugs, and 78 human reported bugs. These
bugs are drawn from the Phase III bugs (see Section~\ref{sec:challengeset}) and
reflect real Google bugs that are not explicitly ruled to be outside
of the scope of Passerine capabilities (e.g., require multimedia) .

For all our experiments, we use Gemini 1.5 Pro (gemini-1.5-pro-001)~\cite{gemini1point5pro} as the LLM underlying Passerine.
We use Gemini out-of-the-box without any additional fine tuning.
LLM calls are performed with temperature = 0.2, and top\_p = 0.95, and we sample the most-likely completion.
Like many generate-and-validate APR approaches~\cite{aprbib}, we perform sampling: the agent is run for 20 (independent) trajectories, where each trajectory proceeds for a maximum of 25 steps. Code search results are limited to 5 file matches, where each 
file match can contain several matching code snippets. As discussed in Section~\ref{sec:commands},
Passerine uses the internal code search, build/test, and filesystem tools available within Google.

\subsection{Patch Correctness}
A critical measure of Passerine performance is its ability
to produce 
both plausible and valid patches. We summarize our findings in Table~\ref{tab:patch-correctness-cost-errors}.
We find that Passerine can effectively generate at
least one valid patch
for a substantial fraction of both machine-reported and
human-reported bugs.
LLM cost has not yet been optimized; we leave this to future work.

\begin{table}[]
\resizebox{\columnwidth}{!}{
\input{tables/patch-correctness-cost-errors}
}
\caption{Summary of correctness and LLM usage for Passerine repairs.
}
\label{tab:patch-correctness-cost-errors}
\end{table}

\begin{figure}
    \centering
    \begin{subfigure}{\columnwidth}
    \centering
    \includegraphics[width=0.8\columnwidth]{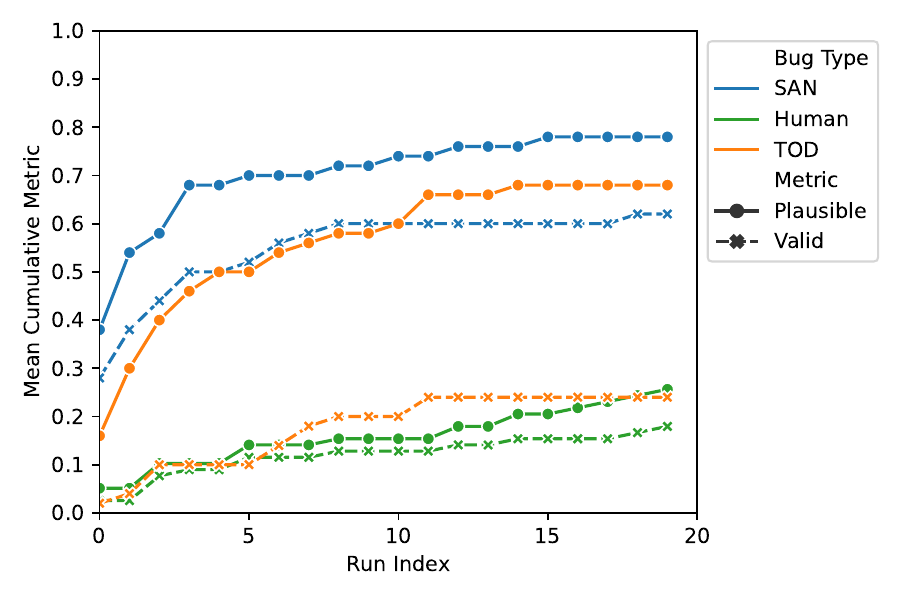}
    \caption{
    Mean cumulative plausible and valid patch generation based on ordered trajectory samples.
    }
    \label{fig:rq2-fail-to-pass-a}
    \end{subfigure}
    ~
    \begin{subfigure}{\columnwidth}
    \centering
    \includegraphics[width=0.8\columnwidth]{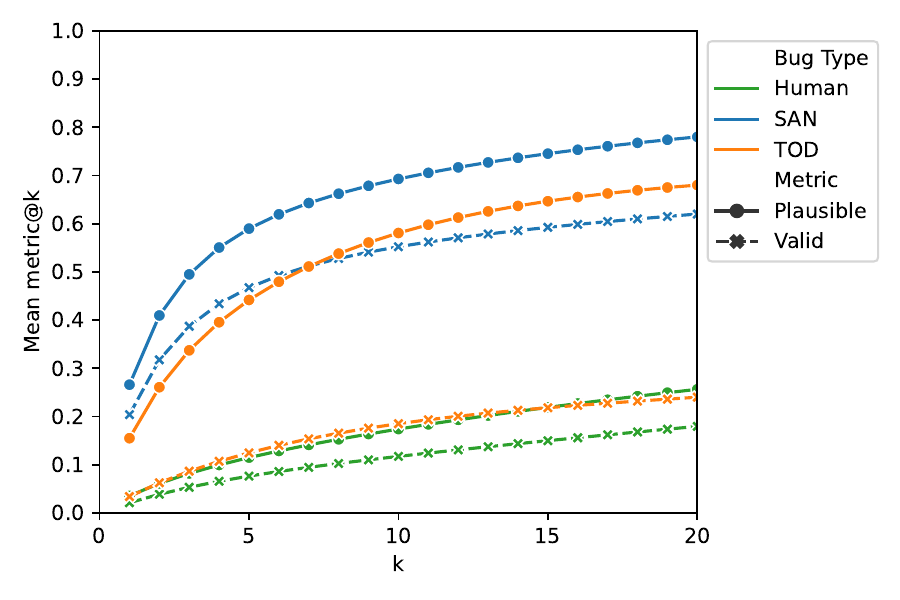}
    \caption{
    Mean plausible@k and valid@k rates, both of which are order-independent and influenced by
    fraction of samples that are plausible or valid, respectively.
    }
    \label{fig:rq2-fail-to-pass-b}
    \end{subfigure}
    \caption{
    Success metrics over 20 trajectories per bug:  Passerine can produce at least one plausible patch for 68\% of TOD bugs, 78\% of SAN bugs, and 25.6\% of Human bugs. Manual annotation shows
   24\% of TOD bugs, 62\% of SAN bugs, and  17.9\% of human-reported bugs have at least one valid patch.
}
    \label{fig:rq2-fail-to-pass}
\end{figure}

Figure~\ref{fig:rq2-fail-to-pass} provides further
detail on patch plausibility
and validity as a function of samples. We observe
that the gap between plausibility and validity varies
by bug type, as a result of nuances in their
testing behavior. TOD bugs, which have the largest
gap, are judged
to be plausible if the test-order dependence is
removed under the original reproduction circumstances,
which specify a single ordering of tests that exhibit
test order dependency; however, this criterion may
be too lenient, and we may filter more solutions by
trying additional test orderings.
For human bugs we find that the opposite is true:
the gap between plausible and valid patch rates is small.
We observe that for these cases the plausibility criterion---the
test contained in the ground-truth commit's test file(s),
which the agent does not have access to---is a strict criterion.

\vspace{-0.2cm}
\begin{center} 
\fbox{
\begin{minipage}[t]{0.97\linewidth}
{\bf Agent-based APR for Google bugs.} 
Our experiments show that Passerine \emph{can}
tackle bugs in Google's enterprise-scale setting, producing
plausible \emph{and} valid patches for both human-reported
and machine-reported bugs. 
\end{minipage}
}
\end{center}

\subsection{Observations}
We now share three key observations from our experiments.

\cheapsubsection{Agent strategies} 
As described in Section~\ref{sec:passerine}, 
we do not provide any high-level guidance (or restriction)
on what commands Passerine can issue or in what order.
As a result of this freedom, we observe that Passerine
can adopt different strategies for different kinds of bug reports.

\begin{figure*}
    \centering
    \begin{subfigure}[b]{0.32\textwidth}
    \includegraphics[width=\textwidth]{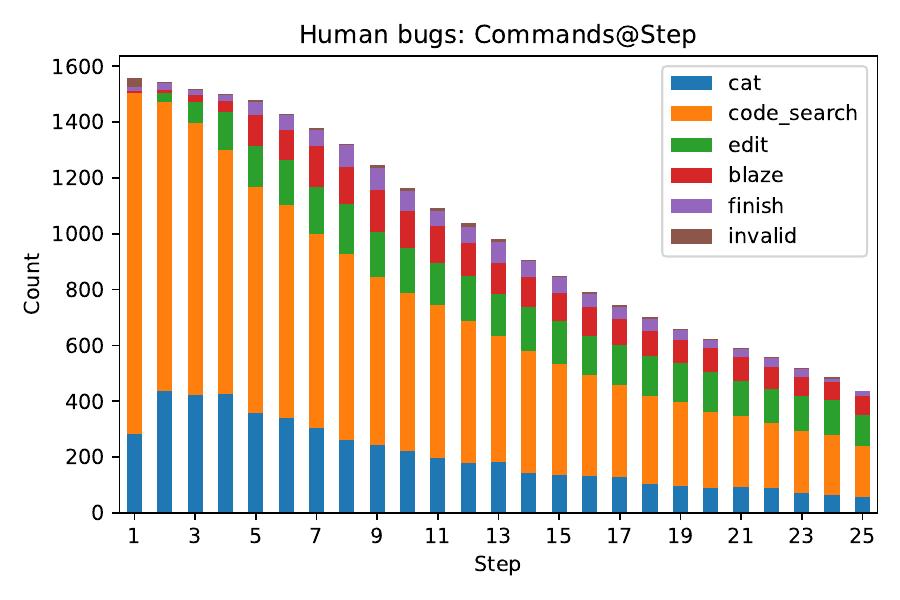}
    \end{subfigure}
    \begin{subfigure}[b]{0.32\textwidth}
    \includegraphics[width=\textwidth]{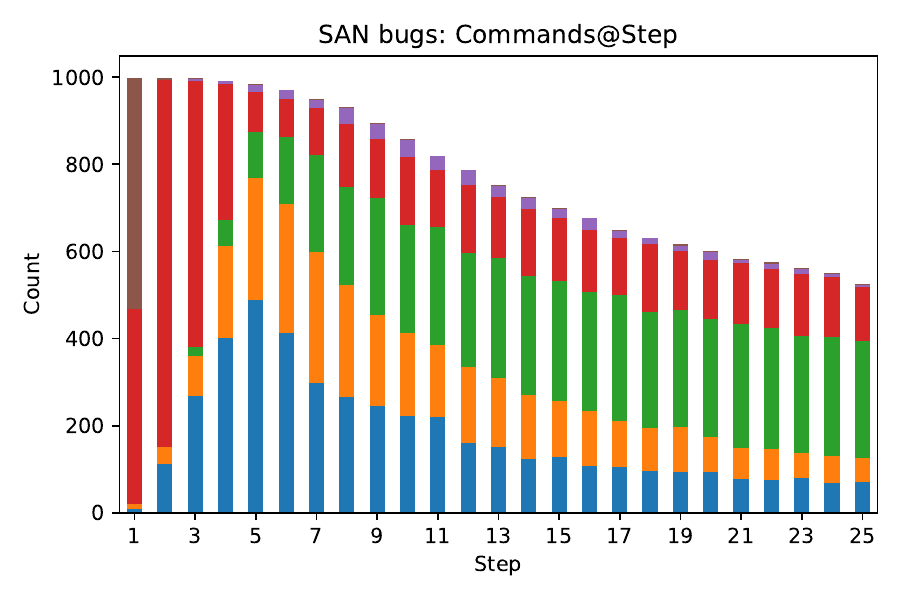}
    \end{subfigure}
    \begin{subfigure}[b]{0.32\textwidth}
    \includegraphics[width=\textwidth]{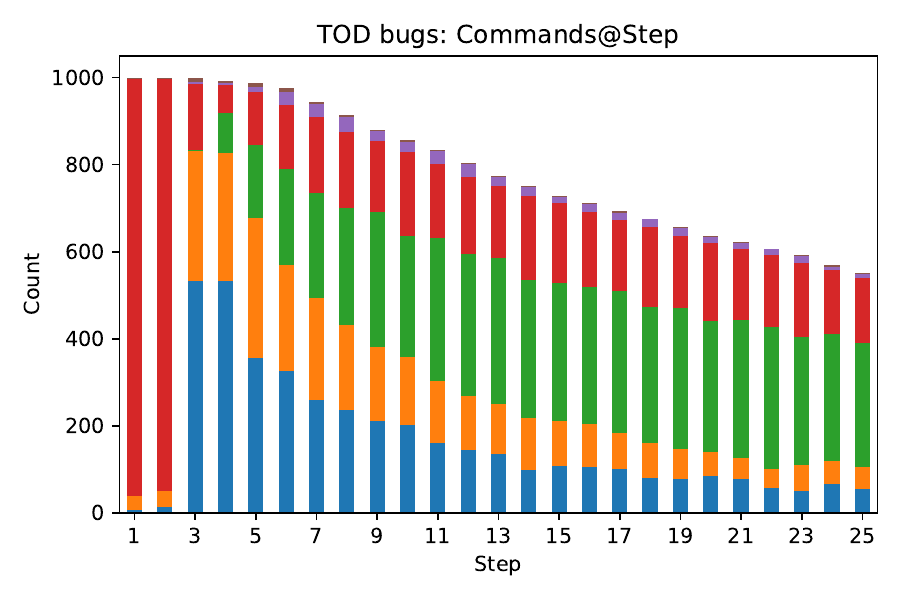}
    \end{subfigure}
    \caption{
    Passerine's trajectory distributions differ substantially when applied to bugs of different types, demonstrating both the distributional differences in the information provided by each bug type and the agent's ability to adapt its behavior to the available information.
    }
    \label{fig:rq2-trajectory-comparison}
\end{figure*}

Figure~\ref{fig:rq2-trajectory-comparison}
shows the frequency of commands by step index in a trajectory, grouped by bug type.
We find that early steps in human-reported bugs are typically dominated
by localization-style operations like 
\lstinline|code_search| and
\lstinline|cat|.
In contrast,
for machine-reported bugs, which, as shown in 
Figure~\ref{fig:bug-reports}, include bug reproduction
information and guidance on the kind of error,
we find that Passerine often starts by running
\lstinline|bazel| (the test building and running command). Later steps in
machine-reported bugs switch to include more edit commands.
Thus, the agent is effectively able to enter a \textit{de facto}
localization when most needed (for human-filed bugs), without
explicitly being constrained to do so, while it can also take
other actions (running tests, for machine-filed bugs) if they
might yield more useful information.

Additionally, we note that for SAN bugs, the first agent step
very often includes an invalid, but unnecessary, command (not in our API)
as a result of the content in the bug description, which
assumes access to \emph{all} standard Google tooling
compared to our restricted set. However, Passerine
adapts quickly and recovers in step 2, where there are no
further such commands.

\vspace{-0.2cm}
\begin{center} 
\fbox{
\begin{minipage}[t]{0.97\linewidth}
{\bf Agent strategy.} 
Despite the lack of high-level strategy guidance (e.g. there is no state machine restricting commands or per-bug-type prompt in our
implementation), Passerine adapts its strategy across bug types.
This same flexibility allows it to recover from incorrect
commands.
\end{minipage}
}
\end{center}

\cheapsubsection{Designing agent-informative bug reports}
To characterize
Passerine's progress on cases where the system \emph{does not}
produce a plausible patch, we consider whether the agent
at least identifies the correct files to edit. Specifically,
we compute the file-system distance between the files edited
by the agent and the ground-truth correct file locations,
where file-system distance is defined as  the distance
between two nodes (files) in an n-ary tree (file system) . 

\begin{figure}
    \centering
    \includegraphics[width=0.7\columnwidth]{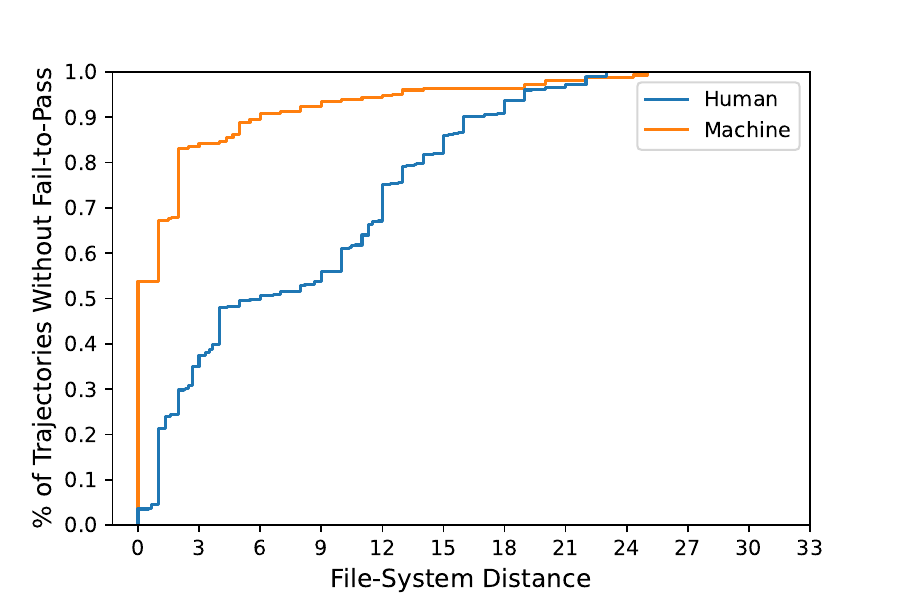}
    \caption{Average file-system distance, defined as distance between two nodes (files) in an n-ary tree (file-system), between the agent edited files and the ground-truth files for trajectories where Passerine \emph{does not produce a fail-to-pass (i.e., not plausible)}.
    Passerine can more effectively localize machine-reported
    bugs as a result of their detailed bug descriptions.}
    \label{fig:rq2-localization}
\end{figure}

Figure~\ref{fig:rq2-localization} shows 
that 53.8\% of machine-reported bug trajectories without a plausible
patch
edited the correct file, compared to only 3.5\% of human-reported bug trajectories without a plausible patch.
So while Passerine cannot produce a plausible patch,
it does a better job at file-level localization for machine-reported
bugs, which have rich bug reports (see Figure~\ref{fig:bug-reports}) that
include description, reproduction information, and test expectations.

The fact that an agent like Passerine can exploit this information
in bug reports, combined with the expectation that
agents will become an increasingly-common part of developer
workflows, leads us to highlight implications for researchers
and practitioners designing bug reporting platforms. Incorporating
nudges for human reporters to enrich reports
with the type of information found in machine-reported bugs
(e.g. bug reproduction guidance) can be a powerful design tool
to increase the effectiveness
of agent-based APR, allowing developers to focus their time on bugs
that are truly out of reach for current agent approaches.

\vspace{-0.2cm}
\begin{center} 
\fbox{
\begin{minipage}[t]{0.97\linewidth}
{\bf Bug reporting.} 
Passerine can produce higher fix rates (or if it fails,
can at least localize the fault to the right file more often) when given
richer bug reports. As agent-based repair continues to gain
traction, the community should consider what nudges can
be provided to human bug reporters to increase the richness
of their issues and put them within reach of existing
agent capabilities.
\end{minipage}
}
\end{center}

\cheapsubsection{Trajectory analysis}
Performing granular inspection of Passerine
trajectories revealed interesting opportunities for additional
optimization. One such opportunity consists of identifying
(and pruning) degenerate agent trajectories, which display
clear misbehaviors. Table~\ref{tab:rq2-smells} shows 
the incidence of two basic properties that signal such
degenerate behavior. 
Trajectories without testing (\texttt{NO\_TEST\_SMELL}) suggest the agent does not
know where to start or how to confirm fixes -- this is observed in 
human bugs.
Interestingly, we also find that more subtle behaviors,
such reading a file that is already in the agent's
context (\texttt{NO\_OP\_CAT\_SMELL}), correlates with failures.
Furthermore, we note that some trajectory dynamics can reflect bug type and difficulty.
Table~\ref{tab:rq2-smells} shows that human bugs are more likely to perform
repeated consecutive searches (\texttt{CONSECUTIVE\_SEARCH}), likely due to
the lack of information contributing to localization challenges as discussed
previously. Both SAN and TOD trajectories, in contrast, are easier to localize
and the agent is more likely to perform repeated edits on the same file (\texttt{CONSECUTIVE\_EDIT}). As might be expected, both of these smells correlate
with bug difficulty and so we observe higher incidence in failing trajectories compared
to trajectories that generate a plausible patch. %

\vspace{-0.2cm}
\begin{center} 
\fbox{
\begin{minipage}[t]{0.97\linewidth}
{\bf Trajectory analysis.} 
Analyzing agent trajectories in detail, such as identifying
repeated misbehaviors, can expose opportunities for optimizations,
some of which, with little effort, could lower reviewer burden.
\end{minipage}
}
\end{center}

\begin{table}[]
\resizebox{\columnwidth}{!}{
\begin{tabular}{lrrrrrr}
\toprule
\multirow{2}{*}{Trajectory Smell} & \multicolumn{2}{c}{Human} & \multicolumn{2}{c}{SAN} & \multicolumn{2}{c}{TOD} \\ \cmidrule(l){2-7} 
 & Failing & Passing & Failing & Passing & Failing & Passing \\
\midrule
NO\_OP\_CAT\_SMELL & 0.44 & 0.22 & 0.70 & 0.53 & 0.60 & 0.54 \\
NO\_TEST\_SMELL & 0.66 & 0.13 & 0.01 & 0.00 & 0.01 & 0.00 \\
CONSECUTIVE\_SEARCH & 0.67 & 0.33 & 0.41 & 0.10 & 0.31 & 0.18 \\
CONSECUTIVE\_EDIT & 0.21 & 0.18 & 0.63 & 0.29 & 0.68 & 0.32 \\
\bottomrule
\end{tabular}
}
\caption{Smell incidence for Passerine trajectories varies across bug type and resulting patch plausibility.}
\label{tab:rq2-smells}
\end{table}

%% file: tables/patch-correctness-cost-errors.tex
\begin{tabular}{@{}llll@{}}
\toprule
                                                     & SAN  & TOD  & Human \\ \midrule
\% bugs with plausible patch                         & 78\% & 68\% & 25.6\%  \\
\% bugs with valid patch                             & 62\% & 24\% & 17.9\%  \\
Avg LLM calls per trajectory                         & 19.4   & 19.8 & 15.89  \\
Avg input characters per LLM call                    & 180K & 136K & 226K \\
Avg output characters per LLM call                   & 418  & 431  & 367   \\
\% trajectories that ended early due to system error & 0.7\%  & 0.2\% & 0.26\% \\ \bottomrule
\end{tabular}%

%% file: discussion.tex
\section{Discussion}

We now discuss ongoing challenges, possible mitigations, and some deployment considerations, all staying within the general framework that we established for Passerine.

\subsection{Challenges and Possible Mitigations}\label{sec:challenges}

\begin{itemize}
    \item Context size: Due to verbose outputs (e.g. test logs, large source code files), and Passerine’s append-only history, the agent’s prompt to the LLM can exceed even large context windows (e.g., 2M tokens).
    We plan to explore possible mitigations including a SWE-Agent-style sliding window which elides the outputs of steps beyond the last n; the use of an LLM to summarize history; and sub-agents with their own, self-contained, history.

    \item Bug reproduction: Machine-reported bugs, like those produced by SAN or TOD, typically include some bug reproduction information. We observe that the agent typically uses this information to reproduce the bug as one of its first steps. In contrast, human-reported bugs include reproduction information at much lower rates. As a result, the agent must first figure out how to expose the bug. We have found that this is itself an interesting area of research, and one that would allow the agent to more effectively perform localization/editing. As a result, we have an ongoing project investigating the ability of an agent to generate bug-reproducing tests.
    
    \item Agent tools: 
    When we inspect agent responses that result in a tool usage failure, we find that for some bugs (particularly
    human bugs), the agent attempts to make use of non-existent commands. One prominent example is
    the agent attempting to use a \texttt{print} command to pose a question to the user
    (despite not having interactive capabilities). For example, in one run Passerine asks
    the user \emph{``print(Please provide more context about the issue with [...] What type of artifacts are missing?''}. Similarly, we observe Passerine can struggle to make progress
    on bugs that would benefit from 
    accessing documents/links listed
    in the bug description and the agent explicitly mentions 
    \emph{``i should try to access it''} (which we do not support).  We plan to extend Passerine's command set
    based on this observed behavior.

    \item Fault localization: We found that for human-reported bugs, fault localization can be challenging -- as reflected by low rates of successful file-level localization
    for trajectories that fail to produce a plausible patch. However, we believe
    there is substantial room for improvement. Currently, Passerine does not explicitly
    task the agent with localization and instead allows the dynamic loop to 
    perform steps as necessary.  In the future, iterative use of code search might be able to provide better localization, as has been argued in the AutoCodeRover~\cite{autocoderover} paper. Furthermore, our
    current implementation of Passerine does not exploit rich development histories
    which may reflect similar past bugs and patches, which can be used
    to further refine possible fault locations~.
    
    \item Diversity of patches: Passerine currently
    samples trajectories independently. Recent work~\cite{chatrepair} showed
    that conditioning LLM generation of patches on earlier generated (and potentially
    validated or rejected) patches can help diversify results in a guided fashion.
    We plan to explore how more sophisticated sampling and search (e.g. beam search)
    can improve Passerine's performance.

\end{itemize}

\subsection{Deployment Considerations}

In an eventual deployment, Passerine may be presented
with bugs that are not within reach of its repair capabilities.
Executing Passerine to attempt to repair such bugs increases compute
costs unnecessarily.
To mitigate this issue, we have begun exploring
\emph{repair abstention}, analogous to abstention in 
classification tasks, to enable Passerine to 
abstain from running on bugs that it is unlikely to fix. 
The filtering criteria defined in Section~\ref{sec:challengeset} can be
used as a guide here, but note that, since the ground-truth patch is not
known for unsolved bugs, the same criteria, which depend on properties of the patch, cannot be applied directly. 
Therefore, applying the same filtering criteria requires auxiliary prediction tasks and knowledge about
how similar bugs were solved in the past.

%% file: threats.tex
\section{Threats and Limitations}
\label{sec:threats}
In this work, we focus on automated repair of bugs drawn from Google's internal repository.
Other industrial settings may have bugs that reflect different characteristics. However,
Google's codebase is likely to share properties with other large software providers.
To create our GITS-Eval dataset we relied on manual assessment of properties that could
not be automatically enforced (see Section~\ref{sec:challengeset}). To mitigate this risk,
we employed multiple annotators.

Importantly, we present Passerine as a proof-of-concept for agent-based APR in an industrial setting.
As a result, we make no claims of how Passerine would fare on other benchmarks such as SWE-Bench. Passerine's
design is also purposefully simple and it is possible improvements in both plausible and valid patch generation
rates would come from a more-sophisticated design. Assessing the viability of such improvements remains part of 
our future work. Finally, we note that patch validity was assessed manually and is subject to the usual
challenges of such evaluation. To mitigate this risk we employed multiple annotators for cases that
were considered ambiguous.

%% file: relatedwork.tex
\section{Related Work}\label{sec:relatedwork}

\subsection{Characterizing bugs and patches in APR
}
Past work has characterized bugs in popular datasets used in automated program repair and more broadly in open source repositories. Sobreira et al~\cite{sobreira2018dissection} characterized the patches for the popular Java-based program repair benchmark. Zhong et al~\cite{zhong2015empirical} explored bugs and their fixes in five large open source projects.  Similar to our GITS/SWE-Bench comparison, these works measured patch size and spread. In contrast to their work, we draw our bugs from GITS, reflecting real tasks for Google developers, and these are multilingual (covering Python, C++, Dart, Go, and Javascript/TypeScript).

\subsection{Agent-based APR}

SWE-Agent~\cite{sweagent}, which directly inspired the implementation of Passerine, implements a ReAct-style loop with access to a set of bash-style APIs as tools. In contrast to the fully dynamic approach present in SWE-Agent, RepairAgent~\cite{repairagent} presents an agent whose actions are constrained by a state machine, disallowing certain tool use sequences. AutoCodeRover~\cite{autocoderover} is an agent-based repair system with tools that exploit explicit program information, such as class/method-based definition search and test-based information for localization. SpecRover~\cite{specrover} augments AutoCodeRover by defining a natural-language specification for the expected behavior at each candidate repair location and introduces a patch review agent. CodeR~\cite{codeR} decomposes APR into multiple subagent tasks that are coordinated with a task graph created and reviewed by a “manager” agent. Similarly, AutoDev~\cite{autodev} employs a multi-agent approach to tackle software engineering tasks beyond program repair, such as test generation and general code completion. MarsCode Agent~\cite{marscode} blends a dynamic, iterative approach to program repair with a traditional generate-and-validate pipeline in a multi-agent repair framework. OpenDevin~\cite{opendevin} — an open-source agent platform inspired by the commercial Devin software engineering agent~\cite{devin} — can be used to build agent-based solutions for tasks in software engineering and other domains.

Similarly to this body of work, Passerine is an agent-based framework. However, our contribution is focused exclusively on program repair. Like many of these approaches, Passerine employs tools in a ReAct style and evaluates its performance using a test-suite-based metric. Our current implementation is aligned with the vision presented in SWE-Agent and exposes a subset of the commands available to that agent. However, in contrast to these systems, Passerine has been developed as a proof-of-concept to tackle bugs in Google’s development environment and as a result has been implemented to use custom tools typically used by Google developers. As a result, our contribution constitutes the first systematic study of  an agent-based APR system in a large industrial codebase.

\subsection{LLM-based APR
}

Prior work has also explored the use of LLMs for fixing and identifying bugs without the need for agents. AlphaRepair~\cite{alpharepair} showed that LLMs could be used to perform zero-shot program repair by framing fixing code as an in-filling task. Agentless~\cite{agentless} obtained results competitive with agent-based approaches on SWE-Bench by employing a fixed repair pipeline, consisting of a hierarchical fault localization approach and multi-patch generation paired with filtering and ranking of patches. ChatRepair~\cite{chatrepair} made use of a conversational-style repair, where the LLM incrementally gains information about patch candidates that fail tests and plausible patch candidates to generate more diverse and accurate patches.
Zhang et al~\cite{yang2024large} also showed that LLMs can be effectively fine-tuned and used as fault localizers for real bugs. 

Like these works, Passerine is also focused on APR, but in contrast it is an agent-based system. Futhermore, Passerine has been designed to tackle bugs within Google’s development environment. Our current evaluation does not use a model fine-tuned for repair nor do we provide meaningful few-shot examples, beyond fixed examples in command APIs to explain how a command is used.

%% file: conclusion.tex
\section{Conclusion}
We present an investigation of the use of an agent-based repair system tailored to Google’s internal software development environment. We collected a benchmark set from Google's internal issue tracking system.
To establish floor for agent-based APR performance on this benchmark, 
we developed Passerine, a minimal agentic repair system with access to Google development tools. We showed that Passerine’s minimal tooling and basic ReAct-style prompting can successfully generate (with 20 trajectory samples) 
plausible patches for 73\% and 25.6\% machine-reported
and human-reported bugs, respectively,
in our evaluation set. Manual inspection showed that 
43\% and 17.9\% of machine-reported bugs 
and human-reported bugs, respectively, had at least one valid patch semantically matching the ground truth. We also find that Passerine automatically adapts its behavior to take advantage of the varying amounts of information provided in machine- and human-reported bugs.
To place these results in context, we also studied how Google's environment can differ from those used in a popular open source repair benchmark, with much larger and spread out patches and with bug descriptions that contain fewer likely code symbols.

\section{Acknowledgements}
We thank Sherry Shi, Sam Cheng, and Alex Polozov for their valuable feedback and assistance.